\def\breakon{\end{multicols}\widetext\vspace{.5cm}
\noindent\rule{.48\linewidth}{.3mm}\rule{.3mm}{.5cm}\vspace{.5cm}}
\def\breakoff{\vspace{.5cm}
\noindent
\rule{.52\linewidth}{.0mm}\rule[-.47cm]{.3mm}{.5cm}\rule{.48\linewidth}{.3mm}
\vspace{.5cm}
\begin{multicols}{2}
\narrowtext}
\documentstyle[aps,multicol,epsf]{revtex}

\newcommand{\ie}{{\it i.e.}}

\begin{document}
\draft
\widetext
\title
{Solitons in Carbon Nanotubes}
\author{Claudio Chamon}
\address{Department of Physics, University of Illinois at Urbana-Champaign,
Urbana, IL 61801-3080
\\
Department of Physics, Boston University,
Boston, MA 02215}
\maketitle
\begin{abstract}
The symmetries of spontaneous lattice distortions in carbon
nanotubes are investigated. 
When the degeneracy of the ground states remains discrete, there are solitons
or domain walls connecting the different symmetry broken vaccua. These
solitons, similarly to the case of polyacetelene, are fractionally charged
states. In addition to the topological domain walls, there are polaron states
with discrete energies within the energy gap. The energies and shapes of
these localized mid-gap states should be accessible via STM spectroscopy.
\end{abstract}

\pacs{PACS:  71.20.Tx 73.20.Dx, 71.45.Lr }

\maketitle
\begin{multicols}{2}
\narrowtext

The electronic properties of carbon nanotubes have recently become the
subject of much attention \cite{American-Scientist}. Single wall
nanotubes, in particular, provide a clean realization of quantum
wires, as well as the opportunity to both engineer electronic device
properties and study fundamental questions in low dimensional
physics. Even within an independent electron approximation, the
properties of the single wall tubes are rather rich and
useful. Depending on how a graphene sheet is wrapped so as to make the
tube, the system can be either an insulator or a metal~\cite{Dresselhaus}.

A great deal of work has been done torwards understanding the role of
electron-electron interactions in the tubes. Studies have been carried
out by using the bosonization scheme
\cite{Dung-Hai,KBF,Egger,Yoshioka}, as well as by mapping the problem
with short range interactions into a two-leg Hubbard model
\cite{BF}. Within the bosonization studies, power law correlations were
found for the order parameter of different electronic instabilities,
such as charge density wave (CDW), spin density wave (SDW) and
superconductivity (SC).

A different perspective is explored in this paper. Here we consider the
effect of lattice deformations in the electronic properties of the nanotubes.
The effects of stress induced long wavelength distortions have been elegantly
studied by Kane and Meele using a tight binding model \cite{KM}. Also,
Peierls like distortions have been previously investigated by studying
displacements along the bond directions, and by assuming no spatial or
quantum fluctuations of these distortions \cite{peierls}. In this paper we
show what exactly the symmetries of the displacement order parameters are,
study the role of fluctuations in this order parameter, and discuss
topological solitons and polarons in carbon nanotubes.

\begin{figure}
\noindent
\hspace{.8 in}
\epsfxsize=2in
\epsfbox{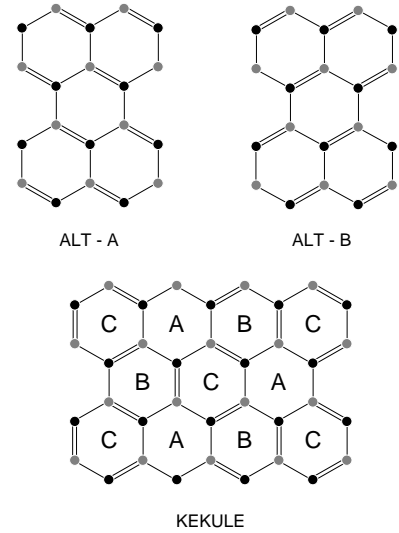}
\vspace{.5cm}
\caption{Dimerization patterns for armchair nanotubes.}
\label{fig2}
\end{figure}

Let us motivate the study of the symmetries of lattice distorsions by
raising a question: are there fractionally charged solitons in carbon
nanotubes? If there is a discrete number of degenerate ground states
corresponding to different lattice distortions, topological
excitations should exist connecting the degenerate vaccua, and
fractionally charged excitations should be present in the domain
walls. These ideas are familiar from another one-dimentional carbon
based system: polyacetylene \cite{SSH}. The carbon nanotubes are
structurally more complex than polyacetelene, and this complexity will
be reflected in the nature of the dimerization patterns that arise
from breaking the lattice symmetries. Consider for example the
patterns shown in Figure \ref{fig2} for armchair nanotubes (the tube
axis is on the horizontal line).

The ALT structures have dimerizations similar to polyacetelene. However,
pairs of rows along the axis have displacements in the opposite direction.
There is a ${\bf Z}_2$ symmetry, and there should be, at the domain walls
between the two-fold vaccua, quantum states with fractional charge $\pm e/2$
{\it per} spin degree of freedom for each of the two species of Dirac
fermions present in the problem \cite{foot1}. The armchair tubes have, in the
low energy spectrum, two species of Dirac fermions, while polyacetelene has
just one. The vaccum flow of charge for the two species of Dirac fermions
goes in opposite directions, and it can be interpreted as flowing charge
$e/2$ from one species to the other. (The precise discussion on the
fractional charge will be done later in the paper).

The Kekul\'e bond-alternated structure contains short (double line)
and long bonds (single line) between neighboring carbon atoms. The
patterned cells are labelled A,B and C according to the relative
position of the single and double lines in the hexagons. If one
visualizes the Kekul\'e structure as a tiling or coloring of the
hexagons in the nanotube with the three labels or colors A,B and C,
there should be three degenerate vaccua corresponding to permutations
of the coloring scheme. 
The domain walls between these three vaccua should be described by
topological solitons with fractional charge $\pm e/3$ {\it per} spin
degree of freedom. 

The conclusions above are based on displacements {\it only} along the
bonds. In order to fully understand the validity of these naive
arguments, we need to look more carefully at more general lattice
distortions and their actual symmetries. We will show that the Kekul\'e
distorsion actually has a continuous $U(1)$ symmetry, and
fluctuations, both thermal and quantum, play an extremely important
role. The ALT structure, on the other hand, is trully two-fold
symmetric.

Let us start from a tight-binding Hamiltonian for a graphite sheet:
\begin{equation}
H=-\sum_{{\bf r}\in {\bf R}}\ \sum_{j=1}^{3} 
\left[ t +\delta t_j({\bf r})\right]
c^\dagger_1({\bf r})\;
c_2({\bf r}+{\bf \tau}_j)
+H.c.\ ,
\label{H-tb}
\end{equation}
where ${\bf r}$ spans the triangular lattice, and the vectors ${\bf
\tau}_j$ ($j=1,2,3$) connect a carbon atom to its three nearest
neighbors in the other sublattice. The distortions of the lattice
alter the bond lengths, and thus the hopping matrix elements change by
$\delta t_j({\bf r})$.

In the absence of the distortions, the spectrum is given by $E({\bf
k})=\pm t\;|h({\bf k})|$, where $h({\bf k})=\sum_{j=1}^3 e^{i{\bf
k}\cdot \tau_j}$. The spectrum contains two Dirac points at ${\bf
K}_\pm=\left(\pm \frac{4\pi}{3a},0\right)$. The dispersion $h({\bf
k})$ can be linearized near the Dirac points, \ie, ${\bf k}={\bf
K}_\pm + {\bf p}$, so the energy near these points is $E({\bf
p})\approx \pm v_F|{\bf p}|$, with a Fermi velocity
$v_F=\frac{3}{2}td$ ($d=a/\sqrt{3}$ is the distance between
neighboring carbon atoms). The nanotubes are obtained from the
graphite sheets by wrapping around a certain direction, identifying
the lattice points $(0,0)$ and $(N,M)$. In the $N=M$ armchair tubes,
the two Dirac points ${\bf K}_\pm$ always lie on the $p_y=0$ sub-band.

{\it The Kekul\'e distortion - } The size of the unit cell is tripled due to
the dimerizations, because the hexagons A,B and C become distinct. This
corresponds to coupling points in the original BZ which are separated by
${\bf G}= {\bf K}_+ - {\bf K}_-$, such as the two Dirac points. 

\begin{figure}
\noindent
\hspace{.3in}
\epsfxsize=2.8in
\epsfbox{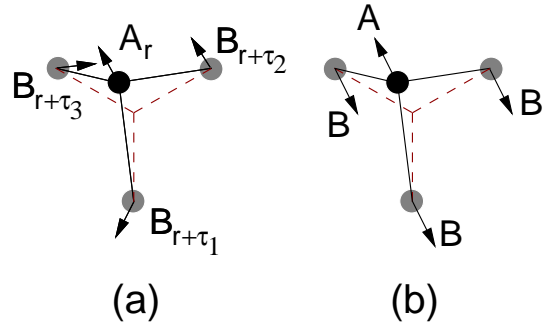}
\vspace{.5cm}
\caption{Displacement vectors for the carbon atoms.
(a) Textured displacements in the Kekul\'e and (b) uniform
displacements in the ALT structures.}
\label{fig4}
\end{figure}

Consider displacements of carbon atoms that can be
written (in terms of the undistorted lattice positions ${\bf r}$) as
$A_{\bf r}= A\;e^{-i{\bf G}\cdot {\bf r}}$ and $B_{\bf r}= B\;e^{i{\bf
G}\cdot {\bf r}}$. Complex numbers are used to represent the
displacement vectors, and $A_{\bf r}$ and $B_{\bf r}$ are in separate
sublattices (see Fig. \ref{fig4}a). Notice that the textures in the
two sublattices spiral in {\it opposite} directions. The Kekul\'e
distortion triples the size of the unit cell, hence the displacements
of the three neighbors to any carbon atom are related by a rotation of
$\pm 2\pi/3$. It is useful at this point to introduce the cubic roots
of unit
$z_j=e^{i{\bf K}_+ \cdot \tau_j}=e^{i \frac{2\pi}{3}(j-1)}
$, and
$
{\bar z_j}=e^{i{\bf K}_- \cdot \tau_j}=e^{-i \frac{2\pi}{3}(j-1)}\ .
$
Is is also convenient to think of ${\bf \tau}_j$ as complex numbers
${\bf \tau}_j=-i\;d\; z_j$. Notice that $\sum_{j=1}^3 z_j=\sum_{j=1}^3
z^2_j=0$ and $\sum_{j=1}^3 z^3_j=3$. In this notation one can write
$B_{{\bf r}+\tau_j}=B\;e^{i{\bf G}\cdot {\bf r}}\;\bar{z}_j$.

The change in bond length $d_j({\bf r})$, at site ${\bf r}$ and in the
direction of ${\bf
\tau}_j$ is
\begin{eqnarray}
\frac{\delta d_j({\bf r})}{d} &=& \left|\frac{\tau_j}{d} 
- \frac{A_{\bf r}}{d} + \frac{B_{{\bf r}+\tau_j}}{d}\right| -1
\nonumber
\\ &\approx& -\frac{1}{2}\;\frac{\bar\tau_j}{d} 
\left(\frac{A_{\bf r}}{d} + \frac{B_{{\bf r}+\tau_j}}{d}
\right)+ H.c.
\nonumber 
\end{eqnarray}

Using the properties of $z_j$, it is simple to show that the
expression above leads to 
\begin{equation}
\frac{\delta d_j({\bf r})}{d} =
i{\bar \epsilon} z_j e^{i{\bf G}\cdot {\bf r}} 
- i{\epsilon} {\bar z}_j e^{-i{\bf G}\cdot {\bf r}}
\label{djd}
\end{equation}
where $\epsilon=\frac{A+ {\bar B}}{2d}$ is the effective lattice
displacement vector that alters bonds. The other combination, namely
$\eta=\frac{A- {\bar B}}{2d}$, changes bond angles without streching
them, only costing elastic energy without any electronic
gain. Therefore, $\eta=0$ or $A={\bar B}$ is chosen.

The elastic energy {\it per} hexagon is
\[
\delta E =\frac{1}{\cal N}\sum_{\bf r}\;\sum_{j=1}^{3}\; 
\frac{1}{2}\;K\;( d_j({\bf r})-d)^2
\]
Using Eq. (\ref{djd}) and the properties of the cubic roots of unit, $z_j$,
one easily finds $\delta E= 3\;K\;d^2 |\epsilon|^2$. This energy cost is
independent of the phase, \ie, the direction of the distortion of the carbon
atoms. This is consistent with a continuous $U(1)$ symmetry, not a discrete
${\bf Z}_3$. Terms that lower the symmetry appear to higher orders in the
expasion of the changes in bond length (as well as bond angle). The
non-linearities, however, are more pronounced in the hopping overlaps, which
are exponentially sensitive to the changes in distance.

Consider a change in bond hopping which is related to the change in
bond length by an exponential: $ t_j({\bf r})=t\;e^{-\alpha\;
\delta d_j({\bf r})/d} $. Expanding to second order, and 
using $d_j({\bf r})$ as given by Eq. (\ref{djd}), one finds
\begin{equation}
\frac{\delta t_j({\bf r})}{t} =
\lambda_j 
\;e^{i{\bf G}\cdot {\bf r}} + {\bar\lambda}_j 
\;e^{-i{\bf G}\cdot {\bf r}} + \alpha^2 |\epsilon|^2\ ,
\label{lambda-r}
\end{equation}
where $\lambda_j =\left(-i \alpha {\bar \epsilon}
-\frac{\alpha^2}{2}\epsilon^2 \right) z_j$.

The wavevector ${\bf G}={\bf K}_+ -{\bf K}_-$ mixes the two species of
Dirac fermions. Substituting Eq. (\ref{lambda-r}) into
Eq. (\ref{H-tb}), one obtains
\begin{equation}
H=\;\sum_{p} \Psi^\dagger(p)\;
\left[
\matrix{{\bf h} & 0 \cr
0 & - {\bf h} \cr
}
\right]
\;\Psi(p)
\ ,
\quad
{\bf h}=\left(
\matrix{ p&\Delta\cr
{\bar \Delta} & - p\cr
}
\right)
\end{equation}
where $\Psi^\dagger=(\psi^\dagger_{+,S}\;
\psi^\dagger_{-,S} \; \psi^\dagger_{+,A} \; \psi^\dagger_{-,A}\;)$, and
\[
\psi_{\pm,S/A}({\bf
p})=\frac{1}{\sqrt{2}}
\;\left[
c_1({\bf K}_\pm + {\bf p})\pm c_2({\bf K}_\pm + {\bf p})
\right]
\]
are the symmetric and antisymmetric linearized fermion operators near
the Dirac points. The order parameter is $\Delta/t
=-3i\;\alpha\;\bar\epsilon +3\alpha^2\;{\epsilon}^2/2$.

The mean field gap $|\Delta|$ that opens is given by $
|\Delta|^2/t^2=9\alpha^2|\epsilon|^2 +9\alpha^3 i\; (\epsilon^3
-\bar\epsilon^3) $, and the cubic terms in $\epsilon$ restore a ${\bf
Z}_3$ symmetry. Notice, however, that these terms are smaller than the
rotational symmetric leading term by a factor of the order
$|\Delta|/t$. Minimizing the elastic and electronic energy for the
filled levels, one finds (ignoring the non-linear effects)
$|\Delta|=v_F \Lambda\; 
\exp\left({-\frac{\pi}{\sqrt{3}\alpha^2}\frac{Kd^2}{t}N}\right)$,
where $v_F \Lambda$ is an energy cut-off scale of the order of the
bandwidth $t$. Using typical parameters for graphite sheets ($t\approx
2.4\; eV$, $K\approx 19.4\; eV/\AA^2$, $\alpha \approx 3.7$, and
$d\approx\;1.42\;\AA$), one finds $|\Delta|\propto t\; e^{-2.1 N}$, so
for a $(5,5)$ tube the gap is of the order $1 K$, as previously found
\cite{peierls}. The anisotropy which restores the ${\bf Z}_3$ symmetry
is a factor $|\Delta|/t$ lower than the gap scale, and it only becomes
apparent at temperatures of the order of $20 \mu K$. This is a very
low scale, and so the symmetry for the Kekul\'e distortion is
effectively $U(1)$.

Even at $T=0$ quantum fluctuations can restore the $U(1)$ symmetry. This can
be studied using a simple rotor model, where the arm of the rotor is the
magnitude of the displacement of the carbon atoms from equilibrium. One finds
that the anisotropy is irrelevant even for small $N$ tubes (the estimated
$N_c$ is less than 2, smaller than that for realistic tubes).
%
%

Phase fluctuations of the order parameter $\Delta(x)$ imply a charge
accumulation $\Delta Q_\pm=\pm\frac{e}{2\pi}\;\Delta \phi$ \cite{Frac-Q},
where $\Delta \phi$ is the phase twist of $\Delta(x)$. The accumulation due
to twisted phases of $\Delta(x)$ coming from the $S,A$ channels have opposite
signs. Notice that these continuous phase twists, and the accompanied charge
compensation between the symmetric and antisymmetric channels, can be
understood in terms of a neutral boson. This is indeed the same situation
that emerges when nearest neighbor electronic interactions are included, and
the system is in the so called CDW2 phase of Krotov, Lee and Louie
\cite{Dung-Hai}. The charge transfered between $S/A$ is not quantized because
the symmetry is a continuous $U(1)$.

{\it The ALT distortion - } We will show that the symmetry for the ALT
distorsion is trully a discrete ${\bf Z}_2$ symmetry. The change in
bond length $d_j$ in the direction of ${\bf \tau}_j$ is now the same
for all lattice points (see Fig. \ref{fig4}b), in contrast to the
textured structure that was treated previously. One has
\begin{equation}
\frac{\delta d_j}{d} =
i z_j {\bar u} - i {\bar z}_j u\ ,
\end{equation}
where $u=\frac{A-B}{2d}$. Analogously to the previous case of lattice
distortion, the elastic energy {\it per} hexagon can be related to
$u$: $\delta E=3\;(K+K_\theta)\;d^2 |u|^2 $, where in this case there
is an extra contribution due to changes in bond angle, as well as bond
length ($K_\theta$ is defined using $d$ to convert from angle to
length displacements - see Ref. \cite{lobo} for values in
graphene). Again, this energy cost is independent of the phase, \ie,
the direction of the distortion of the carbon atoms.

The correction to the Hamiltonian due to the new hopping amplitudes,
however, is not independent of the direction of the displacements. It
is not necessary to keep the changes in bond hopping beyond lowest
order. Similarly to the previous case, one can show that the
Hamiltonian is
\begin{eqnarray}
H&=& v_F\;\sum_{\bf p} \tilde\Psi^\dagger({\bf p})\;
\left[
\matrix{{\bf h}_{A} & 0 \cr
0 & -{\bf h}_{-\bar A} \cr
}
\right]
\;\tilde\Psi({\bf p}) \label{eq:h-alt}
\\
{\rm where}\ \ \ \ 
{\bf h}_{A}&=&\left(
\matrix{0 & p-A \cr
\bar p-\bar A &0 \cr
}
\right)\ ,
\quad
A=2i\alpha\;u/d
\ .
\nonumber
\end{eqnarray}
The spinor $\tilde\Psi^\dagger=(\psi^\dagger_{+,1},
\psi^\dagger_{+,2},\psi^\dagger_{-,1},\psi^\dagger_{-,2})$, 
where $\psi_{\pm,1/2}$ are the fermions near ${\bf K}_\pm$ in the two
sub-lattices.

For the $p_y=0$ band, the distorsion opens the largest gap for real
values of $u$, in which case $A=-\bar A$ and $\Delta = v_F |A| =
3\alpha\; t\; |u| $. There only two vaccua, corresponding to positive
or negative real $u$ (${\bf Z}_2$), as in polyacetelene. Notice that
${\bf h}_{A}={\bf h}_{-\bar A}$, and the spectrum has positive and
negative energies in pairs. If $u$ is purely imaginary, \ie, if the
displacement is orthogonal to the tube axis, then there is no gap and
hence no electronic gain from the negative energy states. There is
only elastic cost for imaginary $u$, so the minimum energy path
connecting the two vaccua should be like in polyacetelene: a real $u$
changes sign.

{\it Quantum numbers for ALT domain walls - } The accumulation of
fractional charge in domain walls between the two-fold vaccua is $e/2$
{\it per} spin degree of freedom, and the $\pm$ species contribute
with opposite phase shifts (one may allow a small imaginary part in
$u$ to see this relative phase), hence the two quantum states have
opposite charge. If filled or occupied, they have charge $\pm e/2$ and
$\mp e/2$ respectively. One can interprete the imbalance as transfer
of charge $e/2$ from one specie of Dirac fermion to the other. 

In the case of polyacetelene, the presence of the two spins ($N_s=2$)
masks the fractionally charged states. Instead, states with quantum
numbers such as charge $e$ and spin $S=0$ appear in the spectrum, as a
consequence of having $N_s=2$ spin species. In the nantotubes, in
addition to two spin states ($N_s=2$) there are two species of
fermions ($N_f=2$) to begin with. 
Because $N_T=N_s N_f=4$, the quantum numbers of the zero energy states in the
nanotubes cannot be distinguisehd from those of electrons. For example, one
can assemble from the fractionally charged states an excitation with charge
$2e$ and spin $S=0$ on the domain wall.

{\it Midgap states and STM probes - } In contrast to polyacetelene chains,
the nanotubes can be individually laid on a substrate, and locally probed via
STM \cite{Liesbeth}. One would then expect that the midgap states with $E=0$
corresponding to domain walls (kinks and anti-kinks) could be probed by
tunneling of electrons from an STM tip. The position dependent tunneling
density of states would probe the shape of the soliton, as well as the energy
of the state.

In addition to the topological zero energy states connecting the two
ground states, there are also polaronic excitations. The difference
between the polarons and the domain walls is that the polarons
correspond to depletions or dimples in the order parameter without
switching between the two ground states. 

It is very simple to obtain the energy levels for the electronic excitations,
as well as the polaron and kink formation energies. We start by recognizing
that the Hamiltonian for the ALT distorsions Eq. (\ref{eq:h-alt}) together
with elastic energy cost $\delta E=3\;(K+K_\theta)\;d^2 |u|^2 $ is simply a
static version of the Gross-Neveu model for a real $u$ background field
\cite{DHN,Campbell,Maki}. The energies of the electronic states and the
formation energies are
\[
\omega_n=\Delta \;\cos\;\left(\frac{n\pi}{2N_T}\right)
\;\; ,
\;\;
E_n=\Delta \frac{2 N_T}{\pi}
\;\sin\;\left(\frac{n\pi}{2N_T}\right)
\]
where $1\le n \le N_T-1=3$ for the polarons, and $n_0=N_T$ for an
infinitely separated kink-anti-kink soliton pair (notice that the
topological electronic state has zero energy). One would hope that
both the topological and the polaronic states could be probed by STM
spectroscopy, with both energy and spatial resolution of the solitonic
states.

{\it Interaction effects - } We would like to briefly discuss the effects of
electron-electron interaction in the above conclusions. The results obtained
in this paper by considering the effects of lattice displacements should be
complementary to the studies of the electron-electron interaction effects
using bosonization (Refs.  \cite{Dung-Hai,KBF,Egger,Yoshioka,BF}). The
interaction effects also open energy gaps, of the same order of magnitude as
the one discussed in this paper (see ref. \cite{Dung-Hai}, for example). The
one major difference between the electron-phonon interaction we discuss in
this paper, and the electron-electron interaction discussed elsewhere, is
that the CDW order parameter for the ALT distortion has a discrete ${\bf
  Z}_2$ symmetry.  Because of the lower symmetry, we expect the soliton and
polaron solutions with mid-gap energies discussed here.

In conclusion, we have studied the symmetries of spontaneous lattice
deformations in carbon nanotubes. We have shown that the Kekul\'e distortion
has a continuous $U(1)$ symmetry, contrary to a naive expectation of a
discrete ${\bf Z}_3$. Consequently, thermal and quantum fluctuations destroy
any long-range order. The ALT structure, however, has a discrete ${\bf Z}_2$
symmetry. We discuss the implications of topological domain walls between the
two-fold ground states, the fractionally charged states on the walls, and the
consequences of having $N_T=N_s\;N_f=4$ in masking the fractional states. We
obtain the energies of the kink states, as well as polaron states, by
recognizing that the Hamiltonian for the ALT distorsion is a version of
Gross-Neveu model with a static real background field. We argue that the
mid-gap states could be probed by STM spectroscopy, resolving experimentally
both the energies and the shapes of the solitonic states.


We would like to acknowledge enlighting discussions with D. K. Campbell, A.
H. Castro-Neto, M. Crommie, E. Fradkin, D.-H. Lee, P. A. Lee, N. Sandler, M.
Stone and L. Venema. This work was supported in part by the NSF through Grant
DMR-94-24511 at the U. of Illinois.

\end{multicols}


\begin{thebibliography}{99}
\bibitem{foot1}
The charge does not scale with the number of rows because only the
$p_y=0$ sub-band is important, and the rows do not behave
independently. What matters is the number of species of fermions.
\end{thebibliography}
\end{document}